\documentstyle[12pt]{article}
\begin{document}
\thispagestyle{empty}
\begin{center}{\Large{On the possibility of detecting torsion from strong gravity, massive neutrino cosmology and PMF}}
\end{center}
\vspace{1.0cm}
\begin{center}
{\large By L.C. Garcia de Andrade\footnote{Departamento de
F\'{\i}sica Te\'{o}rica - IF - UERJ - Rua S\~{a}o Francisco Xavier
524, Rio de Janeiro, RJ, Maracan\~{a}, CEP:20550.
e-mail:garcia@dft.if.uerj.br}}
\end{center}
\begin{abstract}
A cosmological neutrino sea model is used to place bounds on torsion and Lorentz violation and primordial magnetic fields. When one uses gravitational newtonian constant $G_{N}$ we obtain more stringent bounds than the ones obtained by Kostelecky and Russel [PRL,(2001)] which is of the order of $T^{0}\le{10^{-31}GeV}$ axial torsion, which is $T^{0}\le{10^{-38}GeV}$.
When the strong gravity f-meson dominance gravitational constant $G_{f}\sim{10^{38}G_{N}}$ the torsion bound LV is too high of the order of $10^{-2}GeV$. Primordial magnetic field bounds also based on neutrino oscillation is founded to be $B_{\nu}\sim{10^{21}G}$ which is compatible with Enqvist et al [PRL.(1998)]. Earlier the author [MPLA (2011)]has found a bound for torsion of the order $10^{-37}GeV$ from CP ${\alpha}^{2}-dynamos$ which still not as stringent as the present limit of this paper.  By computing the ratio $r=\frac{{\rho}_{B}}{{\rho}_{\gamma}}\sim{10^{-37}}$ in the case of Planck temperatures, shows that r maybe too low to need dynamo action amplify magnetic fields. The most important result of the paper is that making use of cosmological neutrinos in early universe a detectable torsion value of $T\sim{10MeV}$ in present level of energies of LHC is obtained.
\end{abstract}

\newpage\section{Introduction}
Primordial Magnetic Fields are obtained from spontaneous Lorentz invariance broken yielding a photon a mass as was demonstrated by Bertolami and Mota \cite{1}. They were able to place bounds on this field which also yields a galactic dynamo seed with the help of gravitino superpartner of spin-2 graviton from quantum gravity with mass $m_{\frac{3}{2}}\sim{1TeV}$. Here at much lower lever than quantum gravity like $GeV$ of LV in the Lab level \cite{2}. Actually torsion fields as high as $T^{0}\sim{10^{-2}GeV}$ when strong gravity \cite{3} gravitational constant is used. When newtonian gravitational constant is used a torsion bound decays to $\sim{10^{-38}GeV}$. Planck epoch from EC cosmology with spin-torsion high massive neutrino number density in the primordial magnetic field (PMF)of $B\sim{10^{46}G}$ which is much weaker than the magnetic early universe field computed by de Sabbata and Sivaram \cite{3} of $10^{58}G$. We also discuss and compare the effects of these two types of alternative gravities. It is important to note that here we approach the results of Kostelecky of LV in Riemann-Cartan spacetime \cite{4} in SME which is a slight variation of the SM here Higgs bosons and neutrinos fit so well. At $1Mpc$ Bertolami and Mota achieved a result of $10^{-9}G$ in the galactic intracluster which not by coincidence was the same value obtained by Bamba et al \cite{5} using a special case of torsion theory called teleparallelim \cite{6}. It is important to note that the bound we obtained here is different from the one one obtained in the lab cause we are not in laboratory any more. However as far as the LV is concerned there is no problem but certainly if this is also the torsion bound the lower limit from PMF seems not so good cause everybody that works in EC gravity would agree that closer we are from the early universe bigger would be torsion, This let us conclude that strong gravity is really important to deal with the PMFs in the inflationary era for example. Opher and Wichowski \cite{7} have investigated the magnetic fields in Einstein-Cartan gravity but they did not addressed the the problem of galactic dynamo seeds or either the LV and torsion bounds. Other important implication of the
EC theory is in stellar dynamos and gravitational collapse, since according to A Trautman \cite{8} the polarised spins and the interaction with torsion would hold the gravitational collapse so in this framework the EC gravity does not contribute so strong in stars as may contribute to the early universe as could be saw recently in the case \cite{9} of application of spin torsion density of the nucleons orthogonal to a domain wall which gives rise to the PMF and galactic dynamo seeds. In general we need a r coefficient of the order of $\sim{10^{-34}}$ here we found three orders of magnitude less if new data obtained from the paper. Actually pregalactic magnetic fields have $r\sim{10^{-34}}$ and galactic fields $r\sim{1}$ and besides in our case $r\sim{10^{-37}}$ with the torsion effects, this means that torsion effects are not so important in early epochs of universe as far as dynamo mechanism is concerned. Actually Bertolami and Mota have investigate this issue in GR and found that in the range of $10^{-37}\le{r}\le{10^{-5}}$ and suggested that very low limits to r may not render amplification via dynamo process to be effective in the grand unified theory  $M_{GUT}\sim{10^{15}GeV}$. Thus our result for r based on torsion shows that torsion affects dynamo processes in the early universe. The plan of the paper is as follows In section 2 massive neutrinos cosmology comes into play and we compute the cosmological magnetic field associated to these oscillating neutrinos induced by torsion and show that in the case of newtonian gravity no spin-torsion effect is strong enough to contribute to this PMF. In this same section we show how things changes when one introduces strong gravity and section 3 is left for discussions.
 
\section{Torsion bounds from Strong gravity and PMF from massive neutrinos} 
It is now clear that torsion strength is important for its role in the amplification of magnetic fields in the universe it is also clear from the last section that it can play an important role in the early universe when it interacts with spins polarised in the case of stars and nonpolarised in the case of neutrino densities where spatial part of torsion can be neglected and only the time component of torsion is considered as
\begin{equation}
|T^{0}|=\frac{3\pi{G}h}{c^{3}}{\rho}_{\nu}\label{1}
\end{equation}
where the neutrino number density ${\rho}_{\nu}\sim{10^{39}cm^{-3}}$ and we use the newtonian gravitational constant one obtains $|T^{0}|\sim{10^{-38}GeV}$,nevertheless when the strong gravity constant $G_{f}\sim{10^{38} G_{N}}$ the stronger value of torsion obtained is 
\begin{equation}
|T^{0}|\sim{10^{-2}GeV}\sim{10 MeV}
\label{2}
\end{equation}
Now let us compute the neutrino induced PMF with the value of $G_{N}$. To achieve this aim we simply write down the expression for Friedmann equation in EC cosmological with solely the magnetic field energy density on the RHS of the equation 
\begin{equation}
H=\frac{\dot{R}}{R}=\frac{8\pi{G}}{3}[{\rho}-{B}^{2}-\frac{2}{3}\pi{G}\frac{{\sigma}^{2}}{c^{4}}]\label{3}
\end{equation}
where we consider a flat spatial universe $k=0$ and a vanishing cosmological constant $\Lambda$. Since the Hubble factor is always positive or zero in an expanding universe one obtains the following constraint to magnetic energy density
\begin{equation}
B^{2}\le{[{\rho}-\frac{2}{3}\pi{G}\frac{{\sigma}^{2}}{c^{4}}]}\label{4}
\end{equation}
which yields the expression
\begin{equation}
B^{2}\le{[{\rho}_{\nu}-\frac{2}{3}\pi{G_{N}}\frac{{{\sigma}_{\nu}}^{2}}{c^{4}}]}\label{5}
\end{equation}
Taking neutrino spin-torsion density as ${{\sigma}_{\nu}}^{2}\sim{10^{-48} {[\sigma]}^{2}}$ where $[\sigma]\sim{g^{2}dyn^{-1}cm^{-1}}$ is the units and dimension of the spin-torsion ${\sigma}$. Upon computation we note that $B_{\nu}\sim{10^{20}Gauss}$ and the small value of torsion could be useful as an stringent value for LV but certainly would not contribute to the computation of the cosmological magnetic field computed here. let us use the gravitation spin-torsion density at hadronic nuclear matter where $10^{78}$ spin can be polarised on a radius of  which yields a spin-torsion density of $10^{13}cm$ the strong gravity value is used we may recomputed the PMF and obtain the new value
\begin{equation}
B^{2}\le{[{\rho}_{\nu}-\frac{2}{3}\pi{G_{f}}\frac{{{\sigma}_{\nu}}^{2}}{c^{4}}]}\label{6}
\end{equation}
In the case of neutrinos spin-torsion polarised densities are very weak since the neutrino flux are randomic and so a reasonably ${\sigma}^{2}({\nu})\sim{10^{-48}{[\sigma]}^{2}}$ is reasonably and with the neutrino flux of 200 neutrinos one could in strong gravity has a contribution of neutrinos from spin-torsion density as
\begin{equation}
B^{2}({\nu})\le{[{10^{2}}-\frac{2}{3}\pi{10^{30}}\frac{{10^{-48}}}{c^{4}}]}\label{7}
\end{equation}
therefore here even in the case of strong gravity torsion does not contribute to neutrino magnetic fields. Even in the hadron this may not to happen \cite{2}. Let us now see the case where we compute the ratio r between the magnetic energy and the kelvin temperature at neutrino era. In this case $10^{21}Gauss$ as computed above and neutrino temperature is $T_{\nu}\sim{10^{9}K}$ which yields $r\sim{1}$. In the case of hadron era where $B_{h}\sim{10^{19}Gauss}$ and the temperature is $T^{4}\sim{10^{36}K^{4}}$, $r\approx{1}$ which shows that we are in galactic era, but strong gravity was used to obtain this result.
\section{Conclusions}
The main new results of this note is that we may able to obtain more stringent values for the torsion bounds from neutrino cosmological sea as $T\sim{10 MeV}$ which allows one to try the possibility of detecting torsion at the level of energies of experimental particle present levels of energies as LHC and Fermi lab. Previous to the inflation scenario torsion can be strong using strong gravity to think on the possibility of dynamo torsion framework in the early universe dynamo to seed galactic magnetic fields with a very high grand unified mass of GUT. In general it is show that even when strong gravity is used magnetic fields in the post inflationary does not need galactic dynamo from torsion effects. This can be seen from the application of Einstein-Cartan gravity to stellar nuclear matter where gravitational torsion compression actually due to spin-spin contact interaction prevents the presence of a magnetic field and maybe a stellar dynamo without any torsion effect would do the job. Actually was recently shown by the author that as long as a dynamo equation with torsion exists \cite{10} exists galactic dynamo cannot be responsible for amplification field due to the lack of time to the field be amplified before right the galaxy structure formation in the universe. One of the interesting features for future work is to try to get more stringent limits for Lorentz violation in PMF early universe than the one found in reference \cite{11}. Neutrino mass limits are important \cite{12} for future and yet more precise determination of spacetime torsion scales. Further possibilities on torsion detection were discussed in detail by Ilya Shapiro and collaborators \cite{13}.
\section{Acknowledgements}
We would like to express my gratitude to D Sokoloff and 
A Brandenburg for helpful discussions on the problem of dynamos and
torsion and K Subramannian for his kind interest in my work.  Financial support University of State of Rio de Janeiro (UERJ) is grateful
acknowledged.
I would like to express my gratitude to a unknown referee for many important suggestions to improve this paper.


\begin{thebibliography}{12}
\bibitem{1} Bertolami and D F Mota, Primordial magnetic fields via spontaneous Lorentz Breaking, Phys Lett B (1998).
\bibitem{2} N Russel, Lorentz violation and Torsion, Proc of Fourth meeting on CPT Violation, Indiana (2004) world scientific. A Kostelecy Phys Rev D 69, (2004) 085001.
\bibitem{3} V de Sabbata and C Sivaram, Spin and Torsion in Gravitation, (1995) World Sci Publishers. 
L C Garcia de Andrade, Class and Quantum Gravity (2016).
\bibitem{4} A Kostelecy Phys Rev D 69, (2004) 085001.
\bibitem{5} K Bamba, C-Q. Geng, Ling-Wei Luo, JCAP 10 (2012) 058. 
\bibitem{6} A Einstein, Math Annalen Bd 102, 685 (1930).
\bibitem{7} R Opher and U Wichowsi, Phys Rev Lett,
(2013).
\bibitem{8} A Trautman, Nature Phys Sc.242 (1973) 7. W. Kopczynski and A Trautman, Spacetime and Gravitation, Wiley (1992), PWN, Polish Publishers 
\bibitem{9} L C Garcia de Andrade, QCD domain walls in spacetimes with torsion, MPLA (2014). Garcia de Andrade, Mod Phys Lett A,26,11 (2011).
\bibitem{10} L Garcia de Andrade, Phys Lett B 711: 143 (2012). 
\bibitem{11} S Mohanty and U Sarkar, Constraints on background torsion from K-physics, arxiv:hep-th/9804259v1.
\bibitem{12} K Enqvist and H Uibo, Cosmological neutrino mass limits, NORDITA-92/87 P.
K Enqvist, P Olesen and V Semikoz, Phys Rev Lett 69, no 15 (1992) 2157.
\bibitem{13} I Buchbinder, S D Odintsov and I L Shapiro, Effective Action in Quantum Gravity IOP Publishing, (1992).
\end{thebibliography}
\end{document}